\documentclass[ reprint,aps,prl,]{revtex4-1}
\usepackage{amsmath}
\usepackage{graphicx}
\usepackage{dcolumn}
\usepackage{bm}

\begin{document}

\title{Theory of Reciprocating Contact for Viscoelastic Solids}
\author{Carmine Putignano \dag }
\author{Giuseppe Carbone \dag \ddag \S }
\author{Daniele Dini \dag }

\affiliation{\dag  Department of Mechanical Engineering, Imperial College London, London, United Kingdom\\
} 
\affiliation{\ddag Department of Mechanics, Mathematics and Management,
Politecnico di Bari, Bari, Italy }

\affiliation{\S CNR – Institute for Photonics and Nanotechnologies U.O.S.
Bari, Physics Department “M. Merlin”, Bari, Italy. }

\begin{abstract}
A theory of reciprocating contacts for linear viscoelastic materials is
presented. Results are discussed for the case of a rigid sphere sinusoidally
driven in sliding contact with a viscoelastic half-space. Depending on the
size of the contact, the frequency and amplitude of the reciprocating
motion, and on the relaxation time of the viscoelastic body, we establish
that the contact behavior may range from the steady-state viscoelastic
solution, in which traction forces always oppose the direction of the
sliding rigid punch, to a more elaborate trend, never observed before, which
is due to the strong interaction between different regions of the path
covered during the reciprocating motion. Practical implications span a
number of applications, ranging from seismic engineering to biotechnology.

\begin{description}
\item[PACS numbers] 62.20.Qp , 46.55.+d. , 46.35.+z. 
\end{description}
\end{abstract}

\pacs{Valid PACS appear here}
\maketitle

\preprint{APS/123-QED}


\homepage{http://www.Second.institution.edu/~Charlie.Author} 
\affiliation{
 Second institution and/or address\\
 This line break forced}



\section{Introduction}

The mechanics and physics of soft materials are intrinsically complex due to
the strongly time-dependent and usually non-linear constitutive
stress-strain relations that govern their response. Further intricacy is
added when soft bodies are brought into contact and the problem is
exacerbated by the geometry of the intimately mating surfaces. In the last
two decades, the continuously growing technological relevance of engineering
applications involving polymeric materials and soft tissues\emph{\ }has
generated enormous interest in the scientific community and has contributed
to a leap in the number of publications in the field \cite%
{gorb,bio3,vertebrae} ; these span investigations performed across the
scales, from macroscopic to atomistic levels, and include analytical \cite%
{Hunter, Persson2001} numerical \cite{muser,carboneputignano} and
experimental \cite{grosch,triblett} studies. Surprisingly, in spite of these
vast research efforts, our understanding of soft matter problems is
definitely far from being complete.

In this paper, we focus our attention on an issue that has been
systematically ignored but has a crucial importance: the reciprocating
contact of viscoelastic materials, where the relative motion between the
contacting bodies is periodically inverted. Indeed, researchers have almost
universally developed models to investigate unidirectional steady-state
sliding between two mating surfaces made of viscoelastic material \cite%
{Persson2001,carboneputignano}. However, steady-state assumption cannot be
considered a universally valid condition. There is a countless variety of
engineering applications, ranging from the macro- to the nano- scales, where
a periodic inversion of the motion direction is present. Earthquake
viscoelastic dampers are a classic example \cite{earthq1,earthq1}. These
devices are embedded in civil structures to limit the consequences of
earthquakes by introducing a source of damping, that is, beyond the several
possible configurations, the hysteretic dissipation occurring when a set of
rigid punches deforms a layer of rubber.\ Currently, the design in this
field mostly relies on practical and empirical guidelines, and no tool for
quantitative predictions is available. This lack of a robust theoretical
framework involves also very different components, like all the sealing
systems in mechanical applications with an alternate motion \cite{seals}.
Indeed, enhancing performances and efficiency is infeasible without an
accurate knowledge of the interfacial stresses and, consequently, of the
dissipated power. Finally, reciprocating contacts have prominence also at
different scales and in different contexts, like biology and biotechnology (%
\cite{bio1},\cite{bio2}). Skin, ocular system, joints, spine and vertebrae
are some of the examples where viscoelastic soft contact occurs in the human
body. As recently suggested in Ref. \cite{bio1}, this can be observed up to
the cell scale, thus introducing the concept of cell friction. Indeed, Ref. 
\cite{bio1} shows experimental results for reciprocating contact tests on
layers of\ epithelial cells: what is obtained in terms of friction cannot be
explained with a simple elastic model and needs a specific theory.

\begin{figure}[th]
\begin{center}
\includegraphics[width=0.325\textwidth]{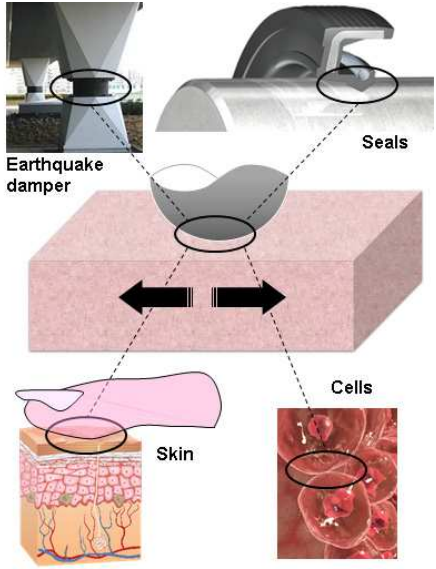}
\end{center}
\caption{Reciprocating viscoelastic contacts: schematic illustration showing
relevant applications.}
\label{Figure0}
\end{figure}

The schematic in Fig. \ref{Figure0} captures the variety of surfaces whose
function and/or performance can be ameliorated by shedding light on the
principles governing the problem under investigation. 

In this work, we develop a theoretical approach and a numerical technique
that, for the first time, allow studying reciprocating contact mechanics
between linearly viscoelastic solids and provide predictions of the response
of the contacting surfaces in terms of stresses, strain and friction. The
paper is outlined as follows. Section II describes the mathematical
formulation which the numerical methodology relies on. Section III focuses
on a simple, tough explicative, case, i.e. the reciprocating sliding contact
of a sphere over a viscoelastic layer. Final remarks are included to comment
on the relevance of the theory and of the results. Appendixes A and B
respectively discuss the validity range of the model and provide a
comparison with steady-state conditions.

\section{Formulation}

The proposed formulation builds on the strengths of the boundary element
method (BEM) in terms of accurately capturing interfacial stresses and
displacements, and, as such, requires the determination of a viscoelastic
reciprocating Green's function $G\left( \mathbf{x},t\right) $.

To this end, let us first assume that the interfacial normal stress
distribution obeys the law $\sigma \left( \mathbf{x},t\right) =\sigma \left[ 
\mathbf{x}-\mathbf{\xi }_{0}\sin \left( \omega t\right) \right] $, i.e. that
the shape of normal stress distribution is fixed but moves on the
viscoelastic half-space with a sinusoidal law of amplitude $\left\vert 
\mathbf{\xi }_{0}\right\vert $ and angular frequency $\omega $. The vector $%
\mathbf{\xi }_{0}\ $also identifies the direction of the reciprocating
motion. Because of linearity and translational invariance, replacing $%
\mathbf{x\rightarrow x}+\mathbf{\xi }_{0}\sin \left( \omega t\right) $
allows to write the relation between interfacial stresses and displacement as%
\begin{equation}
u\left( \mathbf{x},t\right) =\int d^{2}x^{\prime }G\left( \mathbf{x-x}%
^{\prime },t\right) \sigma \left( \mathbf{x}^{\prime }\right) .  \label{eq1}
\end{equation}%
In order to determine $G\left( \mathbf{x},t\right) $, we recall that the
general relation between stress and displacement fields is \cite%
{carboneputignano}:

\begin{align}
u\left( \mathbf{x},t\right) & =\mathcal{J}\left( 0\right) \int
d^{2}x^{\prime }\mathcal{G}\left( \mathbf{x-x}^{\prime }\right) \sigma
\left( \mathbf{x}^{\prime },t\right)  \notag \\
& +\int_{-\infty }^{t}d\tau \mathcal{\dot{J}}\left( t-\tau \right) \int
d^{2}x^{\prime }\mathcal{G}\left( \mathbf{x-x}^{\prime }\right) \sigma
\left( \mathbf{x}^{\prime },\tau \right) ,  \label{visco_principle}
\end{align}%
\newline
where $\mathcal{G}\left( \mathbf{x}\right) $ and $\mathcal{J}\left( t\right) 
$ are the elastic Green's function and the creep material function
respectively. The symbol `$\cdot $' stands for the time derivative. The
creep function is easily linked to the viscoelastic modulus $E(\omega )$ of
the material by means of the relation $1/E(\omega )=i\omega \mathcal{J}%
\left( \omega \right) $ \cite{cristensen}, where $i$ is the imaginary unit
and the Fourier transform of a function $f\left( t\right) $ is $f\left(
\omega \right) =\int dt\exp \left( -i\omega t\right) f\left( t\right) $. The
viscoelastic modulus has the general expression ${1/}E(\omega )=1/E_{\infty
}+\int_{0}^{\infty }d\tau \mathcal{C}\left( \tau \right) /\left( 1+i\omega
\tau \right) $, where $E_{\infty }$ is a real quantity corresponding to the
elastic modulus of the material at very large excitation frequencies. $%
\mathcal{C}\left( \tau \right) >0$ is usually defined as the creep spectrum,
and $\tau $ is the relaxation time \cite{cristensen} . In order to find $%
G\left( \mathbf{x},t\right) $ we choose $\sigma \left( \mathbf{x},t\right) $ 
$=$ $\delta \left[ \mathbf{x}-\mathbf{\xi }_{0}\sin \left( \omega t\right) %
\right] $ and, after substituting in Eq. (\ref{visco_principle}), we obtain%
\begin{align}
G\left( \mathbf{x},t\right) & =\mathcal{J}\left( 0\right) \mathcal{G}\left[ 
\mathbf{x-\xi }_{0}\sin \left( \omega \tau \right) \right]  \notag \\
& +\int_{-\infty }^{t}d\tau \mathcal{\dot{J}}\left( t-\tau \right) \mathcal{G%
}\left[ \mathbf{x-\xi }_{0}\sin \left( \omega \tau \right) \right] .
\label{greenfunct}
\end{align}%
The term $\mathcal{G}\left[ \mathbf{x-\xi }_{0}\sin \left( \omega t\right) %
\right] $ can be re-written as 
\begin{equation}
\mathcal{G}\left[ \mathbf{x-\xi }_{0}\sin \left( \omega t\right) \right]
=\left( 2\pi \right) ^{-2}\int d^{2}q\mathcal{G}\left( \mathbf{q}\right)
e^{-i\mathbf{q}\cdot \left[ \mathbf{x-\xi }_{0}\sin \left( \omega t\right) %
\right] },  \label{elasticGreen}
\end{equation}%
where $\mathcal{G}\left( \mathbf{q}\right) $ is the Fourier transform of the
function $\mathcal{G}\left( \mathbf{x}\right) $ . Now, let us observe that $%
\int d\theta e^{ir\sin \theta }e^{-i\alpha \theta }$ is equal to

\begin{equation}
\int d\theta e^{ir\sin \theta }e^{-i\alpha \theta }=2\pi \sum_{k=-\infty
}^{+\infty }\delta \left( \alpha -k\right) J_{k}\left( r\right) ,
\label{int}
\end{equation}%
where $J_{k}\left( r\right) $ is the $k$-th order Bessel function of the
first kind . Consequently, Eq. (\ref{elasticGreen}) can be casted as%
\begin{equation}
\mathcal{G}\left[ \mathbf{x-\xi }_{0}\sin \left( \omega t\right) \right]
=\sum_{k=-\infty }^{+\infty }A_{k}\left( \mathbf{x}\right) e^{ik\omega t}.
\label{elasticGreen_rewritten}
\end{equation}%
In Eq. (\ref{elasticGreen_rewritten}), $A_{k}\left( \mathbf{x}\right) $ can
be written as

\begin{equation}
A_{k}\left( \mathbf{x}\right) =(2\pi )^{-1}\int_{-1}^{1}ds\mathcal{G}\left( 
\mathbf{x-}s\mathbf{\xi }_{0}\right) B_{k}\left( s\right)
\end{equation}%
with $B_{k}\left( s\right) $ being equal to $B_{k}\left( s\right) =\left(
-i\right) ^{k}T_{k}\left( s\right) B_{0}\left( s\right) $ . $T_{k}\left(
s\right) $ is the Chebyshev polynomial of the first kind and $B_{0}\left(
s\right) =2\left( 1-s^{2}\right) ^{-1/2}$, for $\left\vert s\right\vert \leq
1$ and $0$ otherwise. Substituting (\ref{elasticGreen_rewritten}) in (\ref%
{greenfunct}) we obtain:

\begin{equation}
G\left( \mathbf{x},t\right) =\sum_{k=-\infty }^{+\infty }\frac{A_{k}\left( 
\mathbf{x}\right) }{E\left( k\omega \right) }e^{ik\omega t}.
\label{greeen finale}
\end{equation}%
As mentioned above, the function $G\left( \mathbf{x},t\right) $ has been
obtained under the assumption that the shape of the stress field at the
interface, whose general form is $\sigma \left( \mathbf{x},t\right) =\sigma %
\left[ \mathbf{x}-\mathbf{\xi }_{0}\sin \left( \omega t\right) ,t\right] $,
does not change during the reciprocating motion, \textit{i.e.} $\sigma
\left( \mathbf{x},t\right) =\sigma \left[ \mathbf{x}-\mathbf{\xi }_{0}\sin
\left( \omega t\right) \right] $. Such a condition holds true whenever $%
a_{0}/\left\vert \mathbf{\xi }_{0}\right\vert \ll 1$, where $a_{0}$ the
characteristic dimension of the contact region, and is equivalent to require
that $\left\vert \partial \sigma /\partial t\right\vert /(\left\vert \mathbf{%
\xi }_{0}\cdot \nabla \sigma \right\vert \omega )\ll 1$ (see Appendix A for
more details). This assumption is justified in the majority of cases of
reciprocating contact, and is satisfied point-wise almost everywhere within
the contact area in the analyses presented in this work. Now, to invert the
linear operator in Eq. (\ref{eq1}) we need a numerical approach which
consists in discretizing the contact domain in $M$ square cells. Indeed,
assuming that in each boundary element the normal stress $\sigma $ is
constant and equal to $\sigma _{j}$ , the normal displacement $u_{i}=u\left( 
\mathbf{x}_{i},t\right) $ at the centre $\mathbf{x}_{i}$ of the $i$-th
square can be written as:

\begin{widetext}
\begin{equation}
u_{i}=\frac{1}{N}\sum_{j=1}^{M}\sigma_{j}\sum_{r=1}^{N}L\left[ \mathbf{x}_{i}%
\mathbf{-x}_{j}\mathbf{-}\cos\left( \frac{2r-1}{2N}\pi\right) \mathbf{\xi}%
_{0}\right] \sum_{k=-\infty}^{+\infty}\frac{e^{ik\omega t}\left( -i\right)
^{k}}{E\left( k\omega\right) }\cos\left[ k\left( \frac {2r-1}{2N}\pi\right) %
\right]  \label{eqfinale}
\end{equation}
\end{widetext}

where $L\left( \mathbf{x}\right) $ is related to the Love's solution \cite%
{johnson}. It should be observed that Eq. (\ref{eqfinale}) is obtained by
applying the Chebishev-Gauss quadrature rule to the integral term $%
\int_{-1}^{1}dsT_{k}\left( s\right) L\left[ \mathbf{x}_{i}\mathbf{-x}%
_{j}^{\prime }\mathbf{-}s\mathbf{\xi }_{0}\right] \left( 1-s^{2}\right)
^{-1/2}$ at $M$ nodes, thus making it easier to achieve the numerical
convergence of the problem.

Eq. (\ref{visco_principle}) can be solved by using the iterative technique
developed in Ref. \cite{jmpsel} for elastic contacts, thus providing contact
areas, stresses and strains. It should be noticed that the method does not
require any discretization of the time domain as the time $t$ is treated as
a parameter.

Once the solution is known in terms of stresses and strains, following the
approach stated in Ref. \cite{carboneputignano}, it is straightforward to
calculate the viscoelastic friction force as:

\begin{equation}
F_{T}=\int_{D}d^{2}x\sigma (\mathbf{x})\frac{\partial u}{\partial x}
\label{Friction force}
\end{equation}%
The friction coefficient is then obtained as $\mu =F_{T}/F_{N}$ where $F_{N}$
is the external applied load.

Finally, we conclude noticing that the formulation, in the current form,
does not explicitly account for the role of the tangential tractions at the
contact interface. Indeed, this is out of the scope of our work. The purpose
of the paper is to determine the normal stresses and the normal
displacements distribution, and, on this basis, calculate the viscoelastic
friction that is proportional to the volume of deformed material. It is well
known that the normal and tangential contact problems have only a very weak
coupling, which is normally neglected \cite{Barber}. Furthermore, in the
case of a rigid body in contact with a soft layer --which can be usually
assumed incompressible- , it is absolutely rigorous to assert that tractions
has no influence on normal pressure and displacements, and, consequently, on
the viscoelastic dissipation \cite{Barber}.

\section{Results and discussion}

We study the contact of a rigid sphere of radius $R$ undergoing
reciprocating sliding against a viscoelastic material characterized by one
relaxation time ( being the ratio between the high frequency modulus and the
low frequency $E_{\infty }/E_{0}=11$ and the Poisson ratio $\nu =0.5$). We
assume that the center $\mathbf{x}(t)$ of the sphere moves on the
viscoelastic half-space following the law $\mathbf{x}(t)=[\xi _{0}\sin
\left( \omega t\right) ,0]$. The dimensionless angular frequency of the
reciprocating motion is $\omega \tau $ $=5$ , being $\tau $ the relaxation
time of the viscoelastic material.

\begin{figure}[th]
\begin{center}
\includegraphics[trim=12.5cm 0.5cm 7cm 1.5cm,
clip=true,width=0.615\textwidth,angle=0]{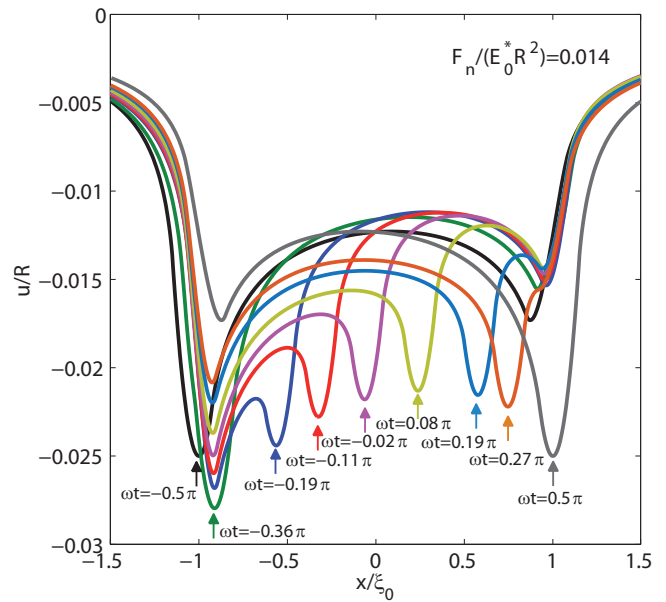}
\end{center}
\caption{The dimensionless normal displacements \ $u(x,y=0)/R$ \ \ as a
function of the dimensionless abscissa $x/\protect\xi _{0}$ for a constant
dimensionless normal force $F_{n}/R^{2}E_{0}^{\ast }=0.014$, for an
amplitude $\protect\xi _{0}/R=1$ and for several values of the dimensionless
time $\protect\omega t\in \left[ -\protect\pi /2,\protect\pi /2\right] $ .}
\label{Figure1}
\end{figure}

Figure \ref{Figure1} shows the evolution of the dimensionless displacements,
\ $u(x)/R$, at the centre of the contact as a function of $x/\xi _{0}$ and
for a specific dimensionless applied normal load $F_{n}/R^{2}E_{0}^{\ast
}=0.014$, and $\xi _{0}/R=1$. Results are shown for different values of $%
\omega t\in \left[ -\pi /2,\pi /2\right] $. An arrow refers, in each case,
to the current position of the sphere. At $\omega t=-\pi /2$ the sphere has
just reached the left dead-point and starts moving from left to right. Upon
reversal of the sliding direction, and for $\omega t\leq -0.36\pi $, a
marked increase of the dimensionless penetration at the center of the sphere
is observed. This is due to the fact that, although the speed is increasing,
it is still too low to cause a significant stiffening of the material, and
the sphere is also moving over a portion of the viscoelastic half-space that
has not yet had the time to relax after the previous contact of the rigid
body. As the sliding speed increases, a non negligible stiffening of the
material and a marked decrease of the penetration are observed (see
displacement in correspondence to the arrow). This is clearly shown by
curves at $\omega t=-0.19\pi ,-0.11\pi ,-0.02\pi $, which also show
additional deformation peaks, one at the left and one to the right of the
arrow: this is the result of the interplay between the deformations, induced
by the indenter as it moves to the right, and the original not yet fully
relaxed footprints left by the sphere at preceding times. For $0<\omega
t<\pi /2$, the sliding speed begins to decrease and the material softens
again, thus leading to an increase of penetration. It is now possible to
justify the occurrence of three different deformations peaks within the
track when the sphere is moving between the two dead-ends: one corresponds
to the current position of the sphere and the other two are located close to
the left and right dead-points respectively, and are the result of the
material inability to fully recover the viscoelastic deformations during a
period of time comparable to the period $T=2\pi /\omega =6.28$ $\mathrm{s}$\
of the reciprocating motion (recall that the relaxation time is $\tau =5~%
\mathrm{s}$).

\begin{figure}[th]
\begin{center}
\includegraphics[trim=0cm 0cm 0cm 0cm,
clip=true,width=0.345\textwidth,angle=0]{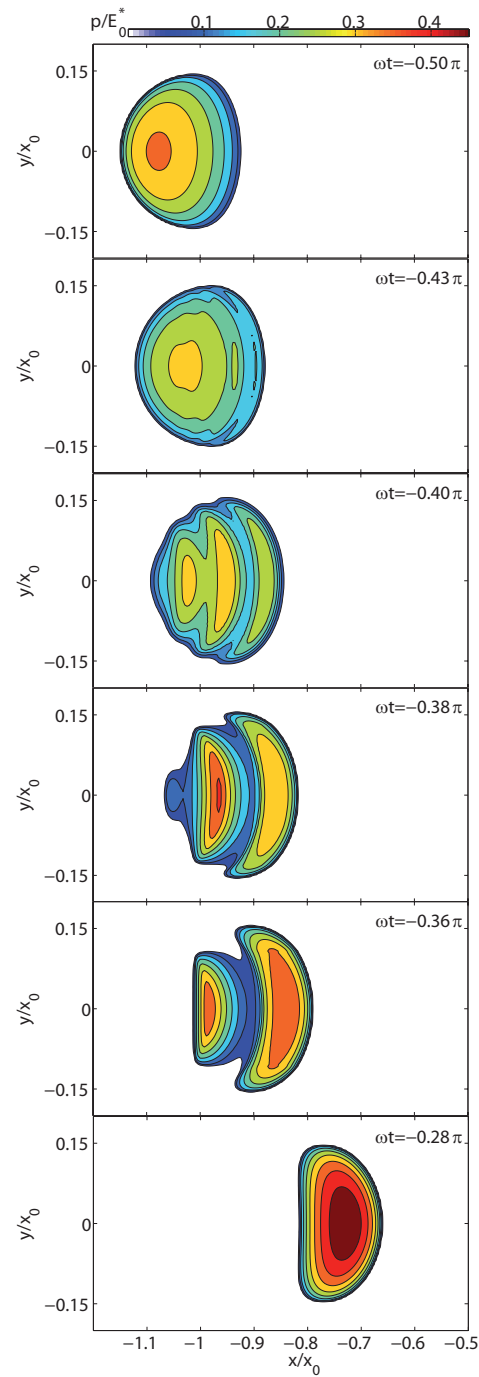}
\end{center}
\caption{The shape of \ the contact area and the contour plots of the
normalised contact pressure distributions, $p/E_{0}^{\ast }$, for several
values of $\protect\omega t$ .}
\label{Figure2}
\end{figure}

The merging or separation of the previous and current sphere footprints,
which takes place close to the dead-points of the reciprocating motion, has
a significant effect on the interfacial normal stress distribution. This is
clearly shown in Figure \ref{Figure2}, which depicts the evolution of the
pressure distribution and shows the shape of the contact area. Let us first
observe that at $\omega t=$ $-\pi /2$, \textit{i.e.} when the sliding speed
goes to zero, the contact area as well as the interfacial normal stress
distribution are characterized by an asymmetric shape. The observed
asymmetry and, in particular, the presence of a peak on the left of the
contact patch is a consequence of the viscoelastic time-delay which prevents
the material to relax immediately when the sliding speed vanishes. As the
sphere starts moving to the right, such a peak cannot disappear suddenly but
has to show a gradual decrease. At the same time, since the punch is
travelling towards the right, as already observed in steady-state
viscoelastic contacts moving at constant velocity \cite{carboneputignano}, a
peak in the pressure distribution has to be originated also at the leading
edge . Finally, at the center of the distribution, where we have the maximum
of the displacement field in the contact area, the pressure must still
resemble the classical elastic Hertzian solution. All this{\normalsize \ }%
process strongly affects the evolution of the pressure distribution at the
interface with the presence of multiple pressure peaks shown by the
snapshots taken at $\omega t=-0.40\pi ,-0.38\pi ,-0.36\pi $ (the reader may
refer to Appendix B to appreciate the difference with steady-state
conditions). A single peaked pressure distribution is later recovered:
indeed, an asymmetric pressure profile marked by a peak closer to the
contact leading edge is visible at $\omega t=$ $-0.28\pi $.

\begin{figure}[th]
\begin{center}
\includegraphics[trim=12.5cm 0.5cm 7cm 1.5cm,
clip=true,width=0.62\textwidth,angle=0]{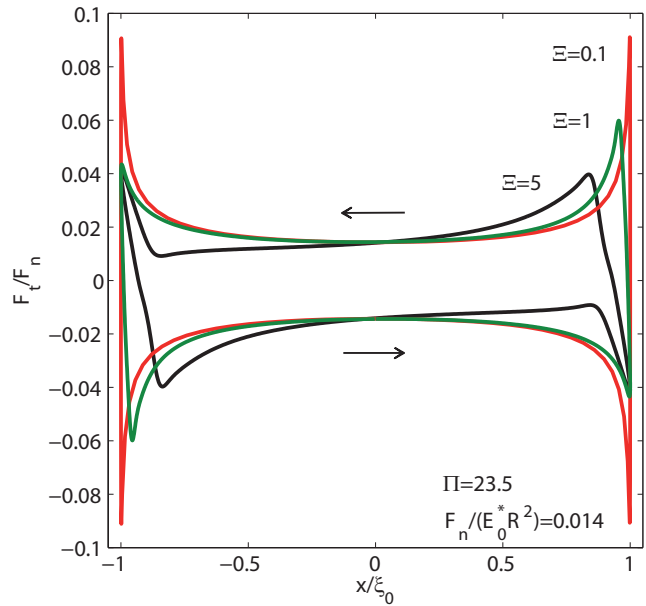}
\end{center}
\caption{The ratio between the tangential and the normal force $F_{t}/F_{n}$
as a function of the dimensionless abscissa $x/\protect\xi _{0}$ for
different values of $\Xi $ . Arrows refer to the hysteresis cycle direction.}
\label{Figure3}
\end{figure}

We may observe that, for a single relaxation time material, in addition to
the ratio $E_{\infty }/E_{0}$, the behavior of the reciprocating contact is
also governed by other two dimensionless parameters. The first dimensionless
group is $\Pi =$ $\tau /t_{0}$, where $t_{0}=a_{0}/\omega \xi _{0}$ and $%
a_{0}\ $is the Hertzian contact radius. This parameter can be also
interpreted as a dimensionless sliding speed \cite{carboneputignano} and
compares the relaxation time $\tau $ with the time $t_{0}$ needed by the
sphere to cover a distance $a_{0}$. The second group, $\Gamma =a_{0}/\xi
_{0}=\omega t_{0}=2\pi t_{0}/T$, compares, instead, the time $t_{0}$ with
the period $T=2\pi /\omega $ of the reciprocating motion. Since we have
earlier assumed that in our problem $\Gamma =a_{0}/\xi _{0}\ll 1$, we can
focus on observing how the solution is affected by $\Pi $. For extremely
small or extremely high values of $\Pi ,$ the response of the system is
elastic (governed by either the high or the low frequency elastic limit of
the material), and no tangential contact force will be generated. At
intermediate values of $\ \Pi $, viscoelasticity will affect the solution
leading to asymmetric contact areas and pressure distributions, and to the
generation of tangential contact forces. In such a case, given the
dimensionless parameter $\Xi =\Gamma ~\Pi =\omega \tau =2\pi \tau /T$, if $%
\Xi <1$, the reciprocating motion will occur on time-scales longer than the
relaxation time $\tau $ of the material and the system will resemble the
steady-state behavior of the contact between a sphere moving on a
viscoelastic half-space at constant speed \cite{carboneputignano}. If $\Xi
\approx 1$, as in the case of Fig. \ref{Figure2}, a strong interaction will
be observed between different viscoelastic regions of the path covered by
the sphere during the reciprocating motion. Note that, under the assumption
small $a_{0}/\xi _{0}$ values (which has been always adopted in this paper),
the condition $\Xi \gg 1$ implies $\Pi \gg 1$, and, in this case, the
elastic response of the material will be recovered: the sphere will be just
performing very fast oscillations, leading to a local stiffening and,
ultimately, to a high frequency elastic behavior.

In Figure \ref{Figure3}, the reduced tangential force, $F_{t}/F_{n}$, easily
calculated once pressures and displacements are known \cite{carboneputignano}%
, is plotted as a function of the dimensionless abscissa $x/\xi _{0}$, which
identifies the position of the sphere along the path, for different values
of $\Xi $ . For $\Xi =0.1$ the material has the possibility to relax before
a single reciprocating cycle is completed. In this case, as the solution
resembles the steady-state viscoelastic sliding contact, the tangential
force $F_{t}/F_{n}$ always opposes the sphere speed at each point along the
path. However, as $\Xi $ is increased (see \textit{e.g.} results for $\Xi =5$
presented in Figure \ref{Figure3}) the relaxation of the material involves
time scales comparable to the time period of the reciprocating motion; in
this case, there exist regions on the sphere track, specifically those close
to the dead-points, where $F_{t}/F_{n}$ has the same direction as the
sliding speed. This is perfectly consistent with the results presented in
Figure \ref{Figure2}.

\section{Conclusion}

This work provides the explicit solution, in terms of a Fourier series, of
the Green's function of the reciprocating contact problem between a rigid
punch and a linear viscoelastic solid. The periodic features, intrinsically
marking the problem, enables the parametric calculation of the contact
solution for each time step without any necessity of employing the solution
in the previous time interval. By implementing such a parametrically
time-dependent approach, we obtain the full numerical convergence in each
moment of the cycle and, interestingly, also when the punch inverts its
motion.

For the simple case of a sphere in contact with a viscoelastic layer, we
show that the behavior of the system is completely determined by two
parameters, that are $\Pi =$ $\tau /t_{0}$ and $\Xi =\omega \tau $ \emph{.}
Depending on these two parameters, the viscoelastic reciprocating contact
may present a wide variety of different behaviors, ranging from the case
that can be captured by the steady-state solution for viscoelastic sliding
to the case where such interactions lead to the formation of a multi-peaked
interfacial pressure distribution and tangential contact forces in (rather
than opposing) the direction of the sliding speed. \ 

This may have important implications in terms of design of materials and
solutions for different applications as it would enable to accurately
capture deformation rates, stress distributions and viscoelastic friction
during reciprocation. Indeed, these are the key quantities to understand
cell growth and skin/counterface interactions, as well as to predict
frictional energy dissipation in \textit{e.g.} mechanical seals and
earthquake dampers.

\appendix

\section{Appendix A: Comments on the parametrically time-dependent Boundary
Element formulation}

When dealing with materials marked by a linear response, the displacement
distribution can be related to the interfacial pressures by means of a
convolution integral with a time- and space-dependent function, i.e. the
Green's function. We may, then, formulate the general contact problem
between a rigid indenter and a linearly viscoelastic slab as:

\begin{equation}
u\left( \mathbf{x},t\right) =\int_{-\infty }^{t}d\tau \int d^{2}x\mathcal{J}%
\left( t-\tau \right) \mathcal{G}\left( \mathbf{x-x}^{\prime }\right) \dot{%
\sigma}\left( \mathbf{x}^{\prime },\tau \right) ,  \label{general formulaion}
\end{equation}

where $\mathbf{x}$ is the in-plane position vector, $t$ is the time, $%
u\left( \mathbf{x},t\right) $ is the normal surface displacement of the
viscoelastic slab, $\sigma \left( \mathbf{x},t\right) $ is the normal
interfacial stress, $\mathcal{G}\left( \mathbf{x}\right) $ and $\mathcal{J}%
\left( t\right) $ are respectively the elastic Green's function and the
creep material function. The relevance of such an approach is related to its
generality: no assumption is made \textit{a priori} on the shape of the
contact domain. Indeed, the method can be employed for any kind of contact
punch and even for rough surfaces: conditions, like periodic boundaries and
finite values of contacting layers thickness, can be easily managed \cite%
{carbone2008,putignanoijss2} . Furthermore, since the creep function $%
\mathcal{J}\left( t\right) $ is absolutely general, the approach is capable
of dealing with any linearly viscoelastic material, from skin tissues to
rubber-based composites.

However, solving directly Eq. (\ref{general formulaion}) may be extremely
challenging: due to the necessity of performing discretization both in time
and space, the computational cost is huge and often infeasible with the
computational technologies currently available. Consequently, when focusing
our attention on the reciprocating contacts, our efforts are aimed at
reducing the computational complexity of Eq. (\ref{general formulaion})
without loosing its generality in terms of contact geometry and material
properties. Indeed, by assuming that the shape of the interfacial normal
stress distribution does not change, i.e. assuming that it obeys the law $%
\sigma \left( \mathbf{x},t\right) =\sigma \left[ \mathbf{x}-\mathbf{\xi }%
_{0}\sin \left( \omega t\right) \right] $ - where $\left\vert \mathbf{\xi }%
_{0}\right\vert $ and $\omega $ are respectively the amplitude and angular
frequency of the sinusoidal law-, Eq. (\ref{general formulaion}) can be
re-written in the form stated in Eq. (\ref{eq1}) . This expression has a
significant advantage: it does not require any discretization of the time
domain since $t$ is present just as a parameter of the viscoelastic
reciprocating Green's function $G\left( \mathbf{x},t\right) $. Such a
formulation enables us to employ the efficient computational techniques
already developed for the purely elastic case \cite{jmpsel} and, therefore,
to find the solution for a reciprocating contact. Incidentally, we observe
that, from a physical point of view, passing from Eq. (\ref{general
formulaion}) to Eq. (\ref{eq1}) is fully justified recalling the periodic
features of the system under investigation.

In this paper, we determine $G\left( \mathbf{x},t\right) $ and investigate
the main peculiarities of the reciprocating contact mechanics. Indeed, when
developing the mathematical formulation, we rely on the aforementioned
condition of a constant shape of the interfacial stress distribution.
Recalling that the total time derivative of pressure field is $\dot{\sigma}%
=\partial \sigma /\partial t+\mathbf{v}\cdot \nabla \sigma $, the condition
implies that, the local rate of change in the pressure $\partial \sigma
/\partial t$, which occurs on time-scales of the order of the period $T$ of
the reciprocating motion, should be negligible compared to the rate of
change of pressure due to the convective term $\mathbf{v}\cdot \nabla \sigma 
$, which occurs on time-scale of $t_{0}$, where $t_{0}$ is the time needed
by the sphere to cover a distance of the order of the contact radius $a_{0}$%
. This then requires that $t_{0}/T\approx a_{0}/\left\vert \mathbf{\xi }%
_{0}\right\vert \ll 1$. This can be easily shown by estimating the local
time derivative and the convective term and requiring that $\frac{\left\vert
\partial \sigma /\partial t\right\vert }{\left\vert \mathbf{v}\cdot \nabla
\sigma \right\vert }\ll 1$, i.e.%
\begin{equation}
\left\vert \frac{\partial \sigma }{\partial t}\right\vert \approx \frac{%
\sigma _{\max }}{T};\text{ }\left\vert \mathbf{v}\cdot \nabla \sigma
\right\vert \approx \frac{\sigma _{\max }}{a_{0}}\omega \left\vert \mathbf{%
\xi }_{0}\right\vert ;\text{ }  \label{estimates}
\end{equation}%
where $\sigma _{\max }$ is the maximum contact pressure and, then, by taking
the ratio of the derivative terms, one obtains:

\begin{equation}
\frac{\left\vert \partial \sigma /\partial t\right\vert }{\left\vert \mathbf{%
v}\cdot \nabla \sigma \right\vert }\approx \frac{\left\vert \partial \sigma
/\partial t\right\vert }{\omega \left\vert \mathbf{\xi }_{0}\cdot \nabla
\sigma \right\vert }\approx \frac{1}{2\pi }\frac{a_{0}}{\left\vert \mathbf{%
\xi }_{0}\right\vert }\ll 1  \label{condition}
\end{equation}

At the end of each stroke during reciprocation, the aforementioned condition
may look critical since the velocity of the sphere tends to vanish; however,
since the time the sphere spends at the dead points of the cyclic sliding
motion is also zero, things have to be observed a bit more carefully.
Indeed, we can calculate the time $t_{0}$ to cover a distance of the order $%
a_{0}$ when the sphere starts moving from the dead point: the distance $a_{0}
$ can be estimated as $a_{0}\approx \left( 1/2\right) \omega ^{2}\xi
_{0}t_{0}^{2}$. One can, then, easily show that $t_{0}/T\approx \left(
a_{0}/\left\vert \mathbf{\xi }_{0}\right\vert \right) ^{0.5}$; hence, if $%
a_{0}/\left\vert \mathbf{\xi }_{0}\right\vert <<1$ , also $t_{0}/T$ will be
sufficiently small to justify the constant shape assumption also at the dead
points.

Incidentally, we observe that we have numerically checked the condition $%
\frac{\left\vert \partial \sigma /\partial t\right\vert }{\left\vert \mathbf{%
v}\cdot \nabla \sigma \right\vert }\ll 1$ for all the cases presented in the
main manuscript once the stress distribution was calculated from the
solution of Eq. (\ref{eq1}).

\appendix

\section{Appendix B: Comparison between steady-state and reciprocating
contacts}

One of the main purposes of this paper is to shed light on the unique
features of the viscoelastic reciprocating contacts. To this aim, it can be
useful here to point out the differences between viscoelastic steady-state
sliding and reciprocating conditions. From a physical point of view, the two
conditions are almost antithetical: in the steady-state case, the punch
always meets underformed material \cite{carboneputignano}; on the contrary,
as we explain in the main manuscript, when dealing with reciprocating
contacts, the rigid punch may deform a region of material that has not yet
relaxed. Only in the limit case of very small values of the parameter $\Xi $
, i.e. given a relaxation time $\tau $ for very small frequencies $\omega $
, the reciprocating case tends to a steady-state-like regime, where the
material has got time to relax before the punch re-engages with it.

\begin{figure}[th]
\begin{center}
\includegraphics[width=0.5\textwidth]{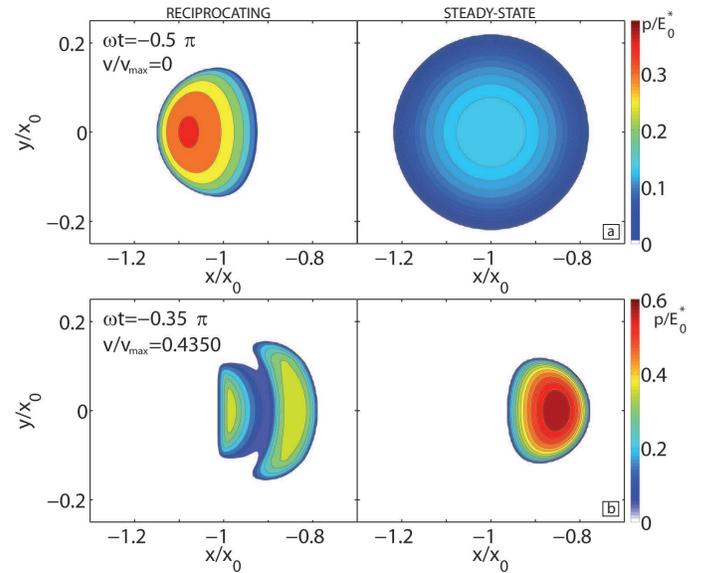}
\end{center}
\caption{The shape of \ the contact area and the contour plots of the
normalised contact pressure distributions, $p/E_{0}^{\ast }$ , in
reciprocating (\textit{on the left}) and steady-state (\textit{on the right}%
) conditions for different values of the time and, consequently, of the
speed. }
\label{figab}
\end{figure}

This physical background entails remarkable differences in terms of
interfacial pressures, normal displacements and, consequently, friction. In
Fig. \ref{figab}, we compare the contour plots of the normal pressure for
the two cases, i.e. the reciprocating contact conditions and the sliding
steady-state contacts. In Fig. (\ref{figab}a) , at the inversion point, i.e.
when the speed is nominally equal to zero, in the reciprocating case, the
pressure still shows a marked asymmetry, that is the consequence of the
viscoelastic time-delay which prevents the material to relax immediately
when the sliding speed vanishes. On the contrary, in steady-state
conditions, the solution at zero speed cannot be anything else that the
elastic classic Hertzian solution with the zero-frequency modulus $E_{0}$.
Furthermore, given a constant normal load, in this last case, due to the
lower modulus, the contact area is much bigger and normal stresses are much
smaller . When the punch starts moving back to the right dead point, we
still have remarkable differences in the pressure distributions (see Fig. (%
\ref{figab}b). 
Such differences are also clearly perceived when looking at the normal
displacements. Upon the motion inversion, we observe the formation of an
additional peak in the displacement distribution: this is impossible in
steady-state conditions. Different distributions in terms of pressure and
normal displacements lead to a different hysteretic curve (see Figure \ref%
{Figure3} in the main manuscript), thus highlighting the importance of the
unique features that characterize reciprocating contact conditions.

\end{document}